\documentclass[useAMS,usenatbib,fleqn,manuscript,usegraphicx]{mn2e}

\usepackage{natbib}
\usepackage{graphicx}
\usepackage{float}
\usepackage{amsmath}
\usepackage[T1]{fontenc}

\title[Newly Discovered RRL Stars]
  {Newly Discovered RR Lyrae Stars in the SDSS$\times$Pan-STARRS1$\times$Catalina Footprint}
\author[M. A. Abbas et al.]
  {M. A. Abbas${^1}$\thanks{Member of the IMPRS for Astronomy \& Cosmic Physics at the University of Heidelberg
and of the Heidelberg Graduate School for Fundamental Physics}
  \thanks{E-mail: mabbas@ari.uni-heidelberg.de},
E. K. Grebel$^{1}$, 
N. F. Martin$^{2,3}$,
W. S. Burgett$^{4}$,
H. Flewelling$^{4}$,\newauthor
R. J. Wainscoat$^{4}$\\
$^{1}$Astronomisches Rechen-Institut, Zentrum f\"{u}r Astronomie der Universit\"{a}t Heidelberg,
M\"{o}nchhofstr. 12--14, D-69120 Heidelberg, Germany \\ $^{2}$Max-Planck-Institut f\"ur Astronomie, K\"onigstuhl 17, D-69117 Heidelberg, Germany\\
$^{3}$Observatoire astronomique de Strasbourg, Universit\'e de Strasbourg, CNRS, UMR 7550, 11 rue de l'Universit\'e, F-67000 Strasbourg, France\\
$^{4}$Institute for Astronomy, University of Hawaii at Manoa, Honolulu, HI 96822, USA}

\begin{document}
\def\citeapos#1{\citeauthor{#1}'s (\citeyear{#1})}

\pagerange{\pageref{firstpage}--\pageref{lastpage}} \pubyear{2013}

\maketitle

\label{firstpage}

\begin{abstract}

We present the detection of 6,371 RR Lyrae (RRL) stars distributed across $\sim$14,000 deg$^2$ of the sky from the combined data of the Sloan Digital Sky Survey (SDSS), the Panoramic Survey Telescope and Rapid Response System 1 (PS1), and the second photometric catalogue from the Catalina Survey (CSDR2), out of these, $\sim$2,021 RRL stars ($\sim$572 RRab and 1,449 RRc) are new discoveries. The RRL stars have heliocentric distances in the 4--28 kpc distance range. RRL-like color cuts from the SDSS and variability cuts from the PS1 are used to cull our candidate list. We then use the CSDR2 multi-epoch data to refine our sample. Periods were measured using the Analysis of Variance technique while the classification process is performed with the Template Fitting Method in addition to the visual inspection of the light curves. A cross-match of our RRL star discoveries with previous published catalogs of RRL stars yield completeness levels of $\sim$50$\%$ for both RRab and RRc stars, and an efficiency of $\sim$99$\%$ and $\sim$87$\%$ for RRab and RRc stars, respectively. We show that our method for selecting RRL stars allows us to recover halo structures. The full lists of all the RRL stars are made publicly available.

\end{abstract}


\section{Introduction}

Studying stars with ages approaching the age of the Universe is of a great importance since they can serve as tracers of the formation and early evolution of galaxies. In particular, they allow us to study the stellar halo which is mainly composed of old stars (e.g. \citealt{johnston2008,schlaufman2009}). It is believed from observations and simulations that mergers and accretions of smaller systems contributed to the formation of the outer halo (e.g. \citealt{bullock2001, bullock2005, carollo2007, mccarthy2012,beers2012}) while the inner halo is a result of accretion of a few massive systems in addition to in situ star formation processes (e.g. \citealt{yanny2003,juric2008,delucia2008,zolotov2010,font2011, schlaufman2012}).

These accretion events and mergers leave signatures in the structure and kinematics of the stellar halo, usually in the form of stellar streams, substructures, and overdensities \citep{ibata1995,newberg2003,duffau2006,schlaufman2009,sesar2010}. It is easier to detect substructures and overdensities of stars at larger Galactocentric distances where the dynamical time scales are longer \citep{bullock2001,bell2008}. If the theoretical picture is correct, we expect an inhomogeneous outer halo that is full of streams and substructures from accreted systems (e.g. \citealt{johnston1998, johnston2008,cooper2010}).


The absence of massive and luminous stars, the old main-sequence turn-offs, the prevalence of horizontal branch stars, and the low metallicities of the halo stars indicate that halo stars are predominantly old. However, it is still unclear whether these stars were mainly formed in situ during the early phase of the collapse of the Milky Way, or whether they were formed outside the Milky Way in satellite galaxies only to be accreted by the Milky Way at a later date (e.g. \citealt{vivas2004, carollo2007, bell2008}). Answers to such questions may be found by identifying and characterizing the streams that the satellite galaxies have left in the halo (e.g. \citealt{zolotov2010}) of the Milky Way where the contamination of foreground stars makes the mapping of stellar structures difficult. Additionally, these streams can serve as very sensitive probes to deduce the shape of the Milky Way's potential \citep{newberg2002,law2009}.

\subsection{RR Lyrae stars as halo tracers}

One way of finding streams is by identifying and mapping RR Lyrae (RRL) stars in the halo. RRL stars are low-mass, core helium burning pulsating stars that fall on the horizontal branch of a stellar population's color-magnitude diagram. RRL stars have a mean absolute $V$-band magnitude of $\langle M_{V} \rangle$ $=$ 0.6 $\pm$ 0.1 \citep{layden1996}, which makes them very good distance indicators. These variable stars are still bright enough to be detected at large distances such as in the halo \citep{ivezic2000}. They have been used as tracers of the chemical and dynamical properties of old stellar populations (e.g. \citealt{kinman2007, bernard2008,keller2008, morrison2009, kinman2009, haschke2012a}) and have served as test objects for theories of the evolution of low-mass stars and for theories of stellar pulsation \citep{smith1995}. 
Many of the substructures that were discovered in the Milky Way were re-confirmed using RRL stars (e.g. \citealt{duffau2006, kepley2007, watkins2009, sesar2010}).

The best and most reliable way to detect RRL stars is by using multi-band time series observations of a sufficiently high cadence and over a sufficiently long period. RRL stars can be divided into fundamental-mode (RRab stars) and first-overtone (RRc stars) pulsators. Since RRL stars are short-period pulsating stars with typical mean periods of $\sim$ 0.57 and $\sim$ 0.34 days for RRab and RRc stars \citep{smith1995}, respectively, the time between observations is preferred to be short in order to sample the magnitudes of the stars at each phase. In addition to that, monitoring an RRL star over a long period of time will result in a more accurate and reliable classification and period determination. Over the past two decades, colors, variability, and light curve properties of RRL stars have been well studied and characterized (e.g. \citealt{smith1995,pojmanski2002, moody2003, vivas2006, wils2006,sesar2010}).

\subsection{Our approach}

In this paper, we use and combine data from different sky surveys out of which each survey has a distinctive advantage that helps in identifying RRL stars with high efficiency (fraction of true RRL stars in the candidate sample), completeness (the fraction of selected RRL stars), and reliability levels. 

First, we apply color cuts to the Sloan Digital Sky Survey (SDSS; \citealt{fukugita1996,york2000,abazajian2009}), 8th data release (DR8; \citealt{aihara2011}). Second, we apply variability cuts using data from the Panoramic Survey Telescope and Rapid Response System 1 3$\pi$ survey (hereafter PS1; \citealt{kaiser2002}) . Finally, we plot light curves and find the periods of the RRL stars using the second photometric catalogue from the Catalina Survey (CSDR2; \citealt{drake2009,drake2013a}) which is based on seven years of multi-epoch observations. Using one of the surveys without the others to find RRL stars results in low efficiency and completeness levels (see Section \ref{usedsurveys}).

In order to define the SDSS color selection threshold limits and the PS1 and CSDR2 variability threshold limits, we use the color and variability properties of the Quasar Equatorial Survey Team (QUEST) RRL star catalogue (QRRL; \citealt{vivas2004,vivas2006}). To compute our efficiency and completeness levels, we compare our results with the RRL stars found in Stripe 82. Stripe 82 ($-50\,^{\circ} \textless $ R.A. $\textless 59\,^{\circ}$, $-1.25\,^{\circ} \textless$ Dec. $\textless 1.25\,^{\circ}$, where both right ascension (R.A.) and declination (Dec.) are given in decimal degrees) covers $\sim$ 270 deg$^2$ of the celestial equator and was observed $\sim$ 80 times by the SDSS. 

\citet{watkins2009} and \citet{sesar2007,sesar2010} independently searched for RRL stars in Stripe 82 using the SDSS data. Because \citeapos{sesar2010} catalog is 100$\%$ efficient and complete, we use it as a test catalog to compute the efficiency and completeness levels of our method.

\subsubsection{Previous Studies}
\citet{drake2013a} had full access to the first photometric catalogue of the Catalina Survey (CSDR1; \citealt{drake2009}), which allowed them to look for RRab stars in the whole $\sim$ 20,000 deg$^2$ area of the sky that was covered by the CSDR1. On the other hand, we had to manually do a multiple object cone search for at most 100 objects at a time as more is not permitted by the public data interface. The CSDR1 RRL star catalogue \citep{drake2013a} contains 12,227 RRab stars found in the CSDR1 \citep{drake2009} database and covers stars with heliocentric distances ($d_{h}$) up to 60 kpc. Because these authors were only interested in RRab stars and to avoid spurious detections, they removed all stars with periods outside the 0.43--0.95 days range. While our paper was nearing completion, \citet{drake2013b} announced the discovery of $\sim$ 2,700 additional RRab stars in a re-analyses of the CSDR1 photometry and using additional data from the CSDR2.

In this paper, we use different variability statistics techniques than the ones used by \citet{drake2013a} and \citet{drake2013b} to find RRab stars. Unlike the latter two studies, we use template fitting techniques to help us in the classification process in addition to the visual inspection of all of the RRL candidate light curves. This allowed us to discover 646 additional RRab stars as compared to \citet{drake2013a} and \citet{drake2013b}. We also found 1,571 RRc stars, of which $\sim$ 1,449 stars are new discoveries.

The properties of the different surveys used in this study are described in Section \ref{usedsurveys}. In Section \ref{identifyrrl}, we describe our method for selecting RRL candidates within the overlapping area between the PS1 and the SDSS. Using the QUEST RRL stars, we define and apply our SDSS color cuts in Section \ref{ColorBoxes} and our PS1 variability cuts in Section \ref{ps1variablity}. In Section \ref{lightcurves}, we use the multi-epoch data from the CSDR2 database to look for stellar variability. The method used to find the periods of the RRL stars is described in Section \ref{aov}. In Section \ref{tfm}, light curves are plotted and the methods used to distinguish RRL from contaminant (non-RRL) stars are described. Section \ref{results} summarizes our results and provides our catalogue of RRL stars. In Section \ref{comp_s82}, we compare our RRL star discoveries with the catalogue of RRL stars in Stripe 82 \citep{sesar2010} to compute the efficiency and completeness levels of our method and periods. The properties of the RRL stars that we missed are also described in the same section. In Section \ref{drake_csdr}, we compare our RRL star discoveries with the catalog of RRab stars from \citet{drake2013a} and \citet{drake2013b} and with the La Silla QUEST (LSQ) catalog of RRL stars \citep{zinn2014}. We discuss our newly discovered RRL stars in Section \ref{newrrs} and we compare them with stars found in the General Catalogue of Variable Stars (GCVS\footnote{Published in 2012 and available from VizieR via http://cdsarc.u-strasbg.fr/viz-bin/Cat?B/gcvs}; \citealt{samus2009}). In Section \ref{substructure}, we find the distances for our RRL stars and we use these distances to recover previously known halo substructures. The results of the paper are summarized in Section \ref{summary}.

\begin{table*}
\centering
\begin{minipage}{900mm}
\caption{The SDSS color cuts.}
\begin{tabular}{@{}llrrrrlrlr@{}}
\hline
$(g-r)_{0}$ $\textless$ 0.4*$(u-g)_{0}$ $-$ 0.16\\
$(g-r)_{0}$ $\textgreater$ 0.4*$(u-g)_{0}$ $-$ 0.67\\
$(g-r)_{0}$ $\textless$ $-$2.5*$(u-g)_{0}$ $+$ 3.42\\
$(g-r)_{0}$ $\textgreater$ $-$2.5*$(u-g)_{0}$ $+$ 2.70\\
$-0.25$ $\textless$ $(g-r)_{0}$ $\textless$ $0.40$ \\
$-0.20$ $\textless$ $(r-i)_{0}$ $\textless$ $0.20$ \\
$-0.30$ $\textless$ $(i-z)_{0}$ $\textless$ $0.30$ \\
\hline
\end{tabular}
\label{colorcuts}
\end{minipage}
\end{table*}

\section{Used Surveys} \label{usedsurveys}
Our method for searching for RRL stars exploits the strong points of different surveys (e.g. SDSS colors of RRL stars) to mitigate other weak points of the same surveys (e.g. lack of SDSS multi-epoch data). As we will show in this study, the synergy between data from different surveys results in an efficient and systematic method to find RRL stars. The different surveys used are described below.

\subsection{The SDSS}

The SDSS \citep{fukugita1996,york2000,abazajian2009} is a photometric and spectroscopic survey containing more than 900,000 galaxies and 110,000 quasars. The SDSS \citep{york2000} uses a 2.5-m telescope located at Apache Point Observatory in New Mexico to image $\sim$ 14,000 deg$^2$ of the sky over a period of 8 years. One part of the SDSS is a multi-wavelength imaging survey (in $u$, $g$, $r$, $i$, and $z$) that goes as deep as 22.3, 23.3, 23.1, 22.3, and 20.8 in $u$, $g$, $r$, $i$, and $z$, respectively \citep{morganson2012}. Most of the SDSS data \citep{stoughton2002, abazajian2009} are based on single epoch observations with the exception of the overlapping regions of adjacent scans and of Stripe 82, which was observed $\sim$ 80 times. 

\citet{watkins2009} searched for RRL stars in Stripe 82 using the public archive of light-motion curves in Stripe 82, published by \citet{bramich2008}. \citeapos{watkins2009} catalog contains 407 RRL stars that lie 5--115 kpc from the Galactic Centre. Additionally, \citet{sesar2010} used the SDSS-Stripe 82 multi-epoch data to discover 483 RRL stars. The light curves, periods, amplitudes, and properties of these stars are all discussed in their paper. \citet{sesar2010} used their discoveries to map the spatial distribution of the halo RRL stars in the 5--120 kpc Galactocentric distance range and were able to detect halo stellar streams and overdensities.

Because \citet{sesar2010} used wider color ranges than the ones used by \citet{watkins2009}, \citeapos{sesar2010} catalog of RRL stars is more complete. According to \citeapos{sesar2010} study, their RRL star discoveries are not contaminated by any other type of stars and are 100$\%$ complete. Thus, we use the latter catalogue in this study to check and test the reliability of our method that is aiming to find RRL stars in the halo.

\subsubsection{The SDSS Observing Technique}

The SDSS telescope uses the drift scanning technique and images the same patch of the sky using its 5 different filters almost simultaneously. This technique yields the true instantaneous colors of the observed sky objects unless they are variable on time scales of less than a few minutes. Since RRL stars have periods between $\sim$ 0.2 and 1 days, the SDSS colors reflect the true colors of the RRL stars. Consequently, since the true colors of RRL stars are available in the SDSS photometric system, we decided to adopt the color cuts of the RRL stars based on the SDSS colors to eliminate most of the non-RRL stars from our sample (see Section \ref{ColorBoxes}).

Contaminant stars that have colors similar to the colors of the RRL stars will still be present in our sample. Potential contaminant stars include non-variable stars (e.g. main-sequence stars with colors at the edge of the color range of the RRL stars), non-RRL variable stars like Ursae Majoris (W UMa) contact binary stars, Algol eclipsing binary stars, $\delta$ Scuti stars, SX Phe stars \citep{palaversa2013}, and stars with large photometric errors. In order to eliminate the non-variable contaminant stars, multi-epoch data are needed. Because most of the SDSS data \citep{stoughton2002, abazajian2009} are based on single epoch observations, we use multi-epoch data from the PS1. 

\subsection{The PS1 3$\pi$ Survey}


The PS1 \citep{kaiser2002} 3$\pi$ Survey began operating from Hawaii in 2010. It uses a 1.8-m telescope that patrols $\sim$ 30,000 deg$^2$ of the sky (north of declination $-30^\circ$) between $\sim$ 10 and 50 times \citep{aller2013} during its three and a half years period of operations. The PS1 uses 5 bandpasses ($g_{P1}$, $r_{P1}$, $i_{P1}$, $z_{P1}$, and $y_{P1}$) that cover the optical and near-infrared spectral range (4,000 \AA $<\lambda<$ 10,500 \AA; \citealt{tonry2012}). At the end of the survey, the PS1 is predicted to go as deep as 23.1, 23.0, 22.7, 21.9, and 20.9 in $g_{P1}$, $r_{P1}$, $i_{P1}$, $z_{P1}$, and $y_{P1}$, respectively, in co-added images \citep{morganson2012}. Individual $g_{P1}$, $r_{P1}$, $i_{P1}$, $z_{P1}$, and $y_{P1}$ exposures that we use in this study have limiting magnitudes of 21.9, 21.8, 21.5, 20.7, and 19.7, respectively \citep{morganson2012}. The PS1 exposure times are filter-dependent and vary between 30s in $z_{P1}$ and $y_{P1}$ and $\sim$ 42s in $g_{P1}$, $r_{P1}$, and $i_{P1}$ \citep{aller2013}. 


\subsubsection{The PS1 3$\pi$ Observing Technique}
Unlike the SDSS, the PS1 images a selected patch of sky using one filter only before moving to the next patch. The PS1 then re-visits the same patch of the sky at a different time to image it using a different filter. This observing technique does not reflect the true colors of the short period variable objects (e.g. RRL stars) as their magnitudes in different filters correspond to different phases. Hence, we favor using the SDSS color cuts to find RRL stars. 

At the same time, the current average number of PS1 clean detections in two (the $g_{P1}$ and $r_{P1}$ bands) out of its five bands are $\sim$ 5 (in each band). We use variability cuts from the PS1 multi-epoch data to distinguish possible variable from non-variable stars of all the stars that pass the SDSS color cuts.

Applying the SDSS color cuts and the PS1 variability cuts does not result in a clean sample of RRL stars. First, variable contaminant stars with colors close to the colors of the RRL stars will still be present. Second, although the PS1 variability cuts will reduce the number of non-variable contaminant stars, they will fail to eliminate all of them because they are based only on $\sim$ 5 epochs in two different filters. With such a small number of epochs, a single outlier can bias the variability statistics. Third, it is not possible to obtain well-sampled light curves and correct periods for RRL stars using only $\sim$ 5 PS1 epochs. Hence, we use the multi-epoch data from the CSDR2 database to study the light curves of our RRL candidates. The CSDR2 light curves allow us to find the periods and subtypes (ab or c) of the RRL candidates very efficiently.

\subsection{The Catalina Survey}

Aiming to discover rare and interesting transient phenomena (e.g. optical transients, Near Earth Objects, etc.), the Catalina Survey \citep{drake2009,drake2013a} uses three different surveys and telescopes: the Catalina Sky Survey (CSS), the Mt. Lemmon Survey (MLS), and the Siding Spring Survey (SSS). While the CSS and MLS are carried out with two different telescopes located in Tuscon, Arizona, the SSS uses a third telescope in Siding Spring, Australia. 

Each telescope is equipped with an unfiltered 4k $\times$ 4k CCD. 2,500 deg$^2$ of the sky are covered by these telescopes every night \citep{drake2013a}. In total, the Catalina Survey telescopes observe around 33,000 deg$^2$ of the sky ($-75\,^{\circ} \textless$ Dec. $\textless 70\,^{\circ}$ and $\mid$b$\mid$ $\textgreater$ 10$^{\circ}$).

Using the SExtractor photometry software, the CSDR2 was released and is now available online\footnote{http://nesssi.cacr.caltech.edu/DataRelease/.}. The CSDR2 contains seven years of observations taken between 2005 and 2011 using the CSS, MLS, and SSS telescopes. The CSDR2 data are available for $\sim$ 500 million objects with $V$-band magnitudes between 11.5 and 21.5 mag.

The CSS uses a 0.7-m Schmidt telescope, which started operating in April 1998 and is located $\sim$ 2,500 meters above the sea level. It uses an unfiltered CCD with a 2$\arcsec$.5 pixel scale providing an 8 deg$^2$ field of view \citep{drake2013a}. The CSS bright and faint magnitude cut-offs are $\sim$ 11.5 and 19.5 mag, respectively. Its typical exposure time is 30 seconds.

The SSS telescope is a 0.5-m Schmidt telescope with a computing system identical to that of the CSS telescope. The SSS telescope began operating in April 2004 and can detect objects with $V$-band magnitudes between $\sim$ 11.5 and 19.0 mag using a CCD camera with a 1$\arcsec$.8 pixel scale and 4.2 deg$^2$ field of view.

The largest telescope used in the Catalina Survey is the MLS telescope. It is a 1.5-m Cassegrain reflector telescope equipped with an unfiltered CCD (1 deg$^2$ field of view and 1$\arcsec$ pixel scale). The MLS can detect objects as bright as $V$ $\sim$ 11.5 mag and as faint as $V$ $\sim$ 21.5 mag, which makes it more sensitive than the CSS and SSS. However, the CSS and SSS cover a much larger area of the sky than the MLS.

The CSDR1 consists of photometry taken by the CSS telescope only and covers an area of $\sim$ 24,000 deg$^2$ of the sky while the CSDR2 consists of photometry taken from all three surveys (CSS, SSS, and MLS) that cover $\sim$ 33,000 deg$^2$ of the sky. The covered areas of the SSS and CSS surveys overlap in the $-20\,^{\circ} \textless$ Dec. $\textless 0\,^{\circ}$ region, while the MLS survey overlaps the CSS and SSS surveys along the ecliptic. It should be kept in mind that the faint limits of the three contributing surveys differ.

Thus, the CSDR2 covers more area than the CSDR1 and at the same time offers a larger number of repeated observations per object when two or more of its three surveys overlap. We use the CSDR2 data in this study to plot the light curves of the RRL candidates in order to correctly classify them and find their periods.

It would have been inefficient to only use the CSDR2 data to find RRL stars in this study because the CSDR2 database allows one to search for only up to 100 point sources in a given query. If we did not apply our SDSS color and PS1 variability cuts we would have had to manually download a large number of data points (mostly for non-RRL stars) from the CSDR2 database before plotting their light curves. Hence, combining data from the SDSS, PS1, and CSDR2 allowed us to find RRL stars in the halo with a higher efficiency level.

\subsection{The Quasar Equatorial Survey Team RRL Stars Survey} \label{questrrl}

QUEST used a 1-m Schmidt telescope located at 3,610 meters elevation at the Llano del Hato Observatory in Venezuela and covered 380 deg$^2$ of the sky \citep{vivas2004}. The QRRL used the QUEST camera, which is a 16 CCDs mosaic with 2,048 $\times$ 2,048 pixels per CCD (1$\arcsec$.02 pixel scale). These CCDs were arranged in a 4 $\times$ 4 array and the entire camera had a field of view of $2^\circ.3$ $\times$ $2^\circ.3$. The QUEST camera allows simultaneous multi-filter photometry as it is designed to operate in a drift-scanning mode. Stars would then pass through the four different filters in the different CCD chips almost simultaneously, similar to the drift-scan imaging technique used in the SDSS. The QRRL is based on $V$-band observations to a limiting magnitude of $V$ $\sim$ 19.5 mag, taken over $\sim$ 2.3 years \citep{vivas2004}. 

The QUEST catalogue of RRL stars contains 498 RRL stars \citep{vivas2004}. According to \citet{vivas2006}, 41 out of the 498 stars are not true variables. These 41 stars were found in crowded regions and the photometric pipeline that was used by QUEST did not include a de-blending algorithm at that time. Consequently, only the remaining 457 stars are used in the analyses in our study.

\subsection{The La Silla QUEST Southern Hemisphere Variability Survey}

The La Silla QUEST (LSQ) Southern Hemisphere Variability Survey used a 1-m Schmidt Telescope at the La Silla Observatory in Chile to observe $\sim$ 1,000 deg$^2$ per night \citep{hadjiyska2012}. It was mainly designed to study supernovae, RRL stars, quasars, and trans-Neptunian objects. The survey is equipped with a broad-band filter (4,000--7,000 \AA) and a camera that consists of 112 CCD detectors. The camera covers an area of $4.\,^{\circ}6 \times 3.\,^{\circ}6$ on the sky. Every one or two days, the LSQ Southern Hemisphere Variability Survey observes the same patch of the sky for 60 seconds twice. These exposures are separated by $\sim$ 2 hours \citep{hadjiyska2012, zinn2014}.

Using its multi-epoch data, \citet{zinn2014} discovered 1,372 RRL stars (1,013 RRab and 359 RRc) with $d_{h}$ in the 5--80 kpc distance range. These stars are distributed across $\sim$ 840 deg$^2$ of the sky in the 150$^{\circ}$--210$^{\circ}$ R.A. and $-10^{\circ}$ to $10^{\circ}$ Dec. range and have been observed between 11 and 300 times.

\begin{figure}
\includegraphics[scale=0.7,angle=-90]{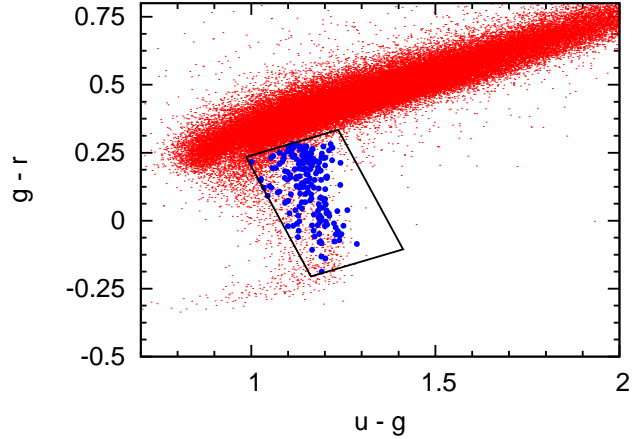}
\caption{The $(u-g)$ $vs.$ $(g-r)$ color-color diagram displaying QRRL stars with DR8 magnitudes in blue dots. The black rhomboidal box indicates our color-color selection cut. Stars that are plotted with red dots and are located inside the black rhomboidal box are considered as RRL candidates and are retained for further analyses. These colors are corrected for the line-of-sight interstellar extinction using the \citet{schlegel1998} dust map. 
\label{colorbox}}
\end{figure}

\section{Identifying RRL stars} \label{identifyrrl} 

We start by using data from the overlapping area between the PS1 and the SDSS that cover $\sim$ 14,000 deg$^2$ of the sky. We first select RRL candidates based on the SDSS $(u-g)$, $(g-r)$, $(r-i)$, and $(i-z)$ colors. As a variability cut, we use the multi-epoch data from the PS1 to distinguish variable from non-variable stars. Applying the SDSS color and PS1 variability cuts in this Section is necessary to reduce the number of light curves to be requested from the CSDR2 as the CSDR2 allows the retrieval of only 100 sky objects at a time.

\subsection{The SDSS Color Cuts} \label{ColorBoxes}
Having different filters is a great advantage as it allows us to construct diagnostic color-color diagrams and consequently helps in having several color constraints for selecting and distinguishing RRL from contaminant stars. Although applying the SDSS color cuts will eliminate a large fraction of contaminant stars, contaminant stars with colors similar to the colors of the RRL stars will still be present (e.g. main-sequence stars with colors at the edge of the color range of the RRL stars, W UMa contact binary stars, Algol eclipsing binary stars, $\delta$ Scuti and SX Phe stars)

We study and identify the $(u-g)$-$(g-r)$ SDSS colors of QUEST RRL stars as this color-color diagram is the most sensitive and efficient in selecting RRL stars. Other single-epoch SDSS colors of RRL stars in the remaining SDSS filters were adopted from \citet{sesar2010}. In order to increase the efficiency of our color cuts, we define and use a $(u-g)$-$(g-r)$ rhomboidal cut instead of a rectangular cut like the one used by \citet{sesar2010}. Additionally, we chose not to use the $(u-g)$-$(g-r)$ colors of the RRL stars found in the catalogue of RRL stars in Stripe 82 \citep{sesar2010} because we wanted to use the latter catalogue as a test catalogue to determine our efficiency and completeness levels, especially because that catalogue is 100$\%$ efficient and complete \citep{sesar2010}. Using the $(u-g)$-$(g-r)$ colors of \citeapos{sesar2010} RRL stars in Stripe 82 would have biased the computation of our efficiency and completeness levels in Section \ref{comp_s82}. At the same time, we cannot accurately compute our efficiency and completeness levels by comparing our results with the QUEST catalogue of RRL stars as the latter catalogue is not complete \citep{vivas2004}.

By positionally cross-matching the 457 QUEST RRL stars with the DR8 database, we obtained the $u$, $g$, $r$, $i$, and $z$ magnitudes of these stars. We used a circle of 3$\arcsec$ centered at the QUEST's positions. Out of the 457 RRL stars, 216 stars are recovered in the DR8 database having $12.0\textless u \textless19.0$. Most of the remaining missed RRL stars have magnitudes beyond our magnitude cut or do not have clean photometry in all of the 5 SDSS filters.

We adopted the above magnitude cut because the PS1 variability of faint ($u\textgreater19.0$) and bright ($12.0\textless u$) stars can be easily biased by the small number of the PS1 repeated observations currently available. Some of the PS1 data are affected by de-blending, cosmic rays, saturation, or non-photometric conditions as the PS1 final calibrated catalogues have not been produced yet. We will investigate much deeper areas of the sky when more PS1 epochs are available and the final PS1 calibrated catalogues are available.

Based on the SDSS colors of the 216 QUEST RRL stars, we define a color-color rhomboidal cut in the $(u-g)$ $vs.$ $(g-r)$ diagram, which is presented in Fig. \ref{colorbox}. These colors are corrected for the line-of-sight interstellar extinction using the \citet{schlegel1998} dust map. We believe that such color corrections can be applied to our stars as these stars are found at high Galactic latitudes in the halo where the overall extinction is small. QRRL stars with DR8 magnitudes are indicated by blue dots and the black rhomboidal box indicates our RRL candidates color-corrected selection box in the $(u-g)$ $vs.$ $(g-r)$ diagram. Only stars that are plotted with red dots and that are found inside the black rhomboidal box are considered as RRL candidates and are retained for further analyses.

Fig. \ref{colorbox} illustrates how RRL stars are distinguished in such a color-color diagram, which demonstrates the usefulness of applying such color cuts. RRL stars follow a trend in this color-color diagram. Their colors always spread out around the blue end of the main stellar locus. The farther away we go from the main stellar locus, the fewer contaminant stars we have. In addition to the $(u-g)$-$(g-r)$ rhomboidal cut that we computed using the 216 QUEST RRL stars, we adopt other SDSS rectangular color cuts from \citet{sesar2010}. All our color cuts are listed in Table \ref{colorcuts}.

Since we use the catalogue of RRL stars in Stripe 82 \citep{sesar2010} as a reference catalogue to compute our completeness and efficiency levels in Section \ref{comp_s82}, computing our own $(u-g)$-$(g-r)$ SDSS color cut using the QUEST RRL stars and then applying them to the stars found in Stripe 82 ensures that the completeness and efficiency levels we achieve are unbiased. Additionally, \citet{sesar2010} use rectangular cuts while we use a $(u-g)$-$(g-r)$ rhomboidal cut. The SDSS $(u-g)$ color serves as a surface gravity indicator for these stars. The range ($\sim$ 0.3 mag) and the $rms$ scatter ($\sim$ 0.06 mag) are the smallest in this color \citep{ivezic2005}.

Within the overlapping area between the PS1 and DR8, 308,342 stars passed all the color cuts listed in Table \ref{colorcuts}.

\subsection{The PS1 Variability Cuts} \label{ps1variablity}

Because many stars passed the SDSS color cuts and because the CSDR2 allows us to do a manual search of only 100 sky objects at a time, we use the PS1 preliminary variability cuts to reduce the number of RRL candidates light curves to be retrieved from the CSDR2.

On average and at the end of the survey, the number of the PS1 pointings per object will be around 12 per filter. Taking chip gaps and dead cells into account, the PS1 observations per object will likely amount to about 9 times in each filter. Since this survey is still continuing, the average number of clean detections in the $g_{P1}$ and $r_{P1}$ filters are currently only $\sim$ 5. These are the ``good'' detections, which means that these detections were not saturated or blended, and were not flagged as cosmic rays \citep{morganson2012}. We will include more epochs from the PS1 in future studies when they are available.

The PS1 photometric catalogue contains the average $g_{P1}$, $r_{P1}$, $i_{P1}$, $z_{P1}$, and $y_{P1}$ magnitudes for each point source that were computed using the multi-epoch data available. It also contains the standard deviations ($\sigma$) that show the scatter of the single-epoch data about the average magnitudes in each filter. As a preliminary variability cut, we select stars with a standard deviation greater than 0.05 in $g_{P1}$ ($\sigma_{g_{P1}}$) or in $r_{P1}$ ($\sigma_{r_{P1}}$). The mean $g_{P1}$ and $r_{P1}$ errors are $\sim$ 0.02 mag in both filters. Although this preliminary variability cut does not result in a clean sample of variable stars, it eliminates a large fraction of non-variable contaminant stars (e.g. main-sequence stars) and reduces the number of CSDR2 light curves to be requested. The PS1 variability statistics are based on $\sim$ 5 epochs in two different filters where a single outlier can bias the statistics. Potential contaminant stars that can still be present after the preliminary variability cut include non-variable stars (e.g. main-sequence stars) with relatively large $g_{P1}$ or $r_{P1}$ photometric errors and non-RRL variable stars with colors similar to the colors of RRL stars.

Around 34,200 stars (11$\%$ of the 308,342 stars selected in Section \ref{ColorBoxes}) passed these two variability selection cuts. Most of the stars that passed our SDSS color cuts but not the PS1 variability cuts are main-sequence stars with colors close to the edge of the color range of the RRL stars. We use the CSDR2 multi-epoch data in the next section to obtain a clean list of variable stars and to carry out a light curve analysis.

\section{The CSDR2 Light-Curves} \label{lightcurves}
\begin{figure}
\centering
\includegraphics[scale=0.58,angle=-90]{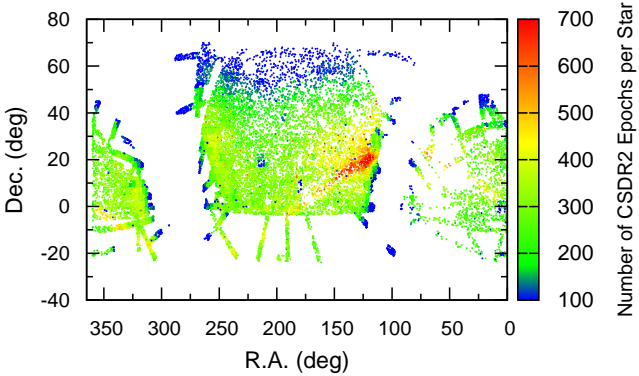}
\caption{The number of CSDR2 observations per star as a function of equatorial J2000.0 right ascension and declination. These are only the stars that are in the footprint of the SDSS covering $\sim$ 14,000 deg$^2$ of the sky. The values are color-coded according to the legend. The mean number of epochs of these stars in the CSDR2 database is 270. 90$\%$ of these stars were visited more than 100 times. 
\label{distribution_rr}}
\end{figure}

We extract the CSDR2 light curves for the stars that passed the SDSS color and the PS1 variability cuts. We searched for all of our $\sim$ 34,200 RRL candidates and found $\sim$ 21,050 stars in the CSDR2 database. The remaining stars were either not observed with the Catalina Survey, or were found in crowded regions. We used a circle of 3$\arcsec$ centered at the DR8 positions. The mean number of epochs for these stars in the CSDR2 database is 270. More than 90$\%$ of these stars were observed more than 100 times. The number of CSDR2 observations per star as a function of equatorial J2000.0 R.A. and Dec. is illustrated in Fig. \ref{distribution_rr} where the values are color-coded according to the legend. Because 90$\%$ of the stars have more than 100 epochs, the CSDR2 variability statistics and light curve analyses were sufficient and very accurate to reliably distinguish RRL from contaminant stars.

In order to get rid of possible outliers, we omit data points that are more than 3$\sigma$ from the CSDR2 mean magnitudes ($Mag$) for each star. This step ensures a reliable variability statistics and better phased light curves (light curves folded using a specific period). This is because some of the CSDR2 observations were taken under non-photometric conditions and sometimes possibly due to the recalibration process itself. Potential outliers can also be contaminated by cosmic rays.

Using data from the CSDR2, we calculated variability statistics like the weighted standard deviation ($W_{\sigma}$), $\chi^2$, variability index ($\alpha$), variance ($Var$), amplitude ($Amp$), and the skewness of the light curves ($\gamma$) for a better separation of variable and non-variable stars. We list the definitions of these quantities below.

\begin{equation}  \label{WeightedMeanMag}
W_{<m>} =  \frac{\sum\limits_{i=1}^N \frac {x_i}{x_{ierr}^2} }{ \sum\limits_{i=1}^N \frac {1} {x_{ierr}^2}}
\end{equation}

\begin{equation} \label{Wstd}
W_{\sigma} =  \sqrt {\frac{\sum\limits_{i=1}^N      \frac {(x_i-W_{<m>})^2}{x_{ierr}^2}}{  \frac {N-1} {N  } \sum\limits_{i=1}^N  \frac {1} {x_{ierr}^2  } }}
\end{equation}

\begin{equation} \label{chi}
\chi^2 =\frac{1}{N-1} \sum\limits_{i=1}^N \frac{ (x_i-\bar{x})^2} {x_{ierr}^2 }
\end{equation}

\begin{equation} \label{equation_vindex}
\alpha = \frac{\sum\limits_{i=1}^N ( x_i-\bar{x})^2-x_{i_{err}} }{N-1}
\end{equation}

\begin{equation} \label{skew}
\gamma = \frac{N}{(N-1)(N-2)}\frac{1}{\zeta^3} \sum\limits_{i=1}^N  (x_i-\bar{x})^3 
\end{equation}

\begin{equation} \label{zeta}
\zeta = \sqrt {  \frac{1}{N-1}\sum\limits_{i=1}^N (x_i-\bar{x})^2}
\end{equation}

\noindent
where $x_i$ and $x_{i_{err}}$ represent the CSDR2 single-epoch magnitude and the error corresponding to this magnitude, respectively. $\bar{x}$ is the mean magnitude ($Mag$), and $N$ is the number of the CSDR2 epochs for each star. 

Our RRL candidates are the stars that passed our SDSS color cuts in Section \ref{ColorBoxes}, the PS1 variability cuts in Section \ref{ps1variablity}, and that have $W_{\sigma}$, $\chi^2$, $\alpha$, $Var$, and $Amp$ greater than 0.1, 1.0, 0.002, 0.006, and 0.4, respectively, and $-1.0 \textless \gamma \textless 1.0$. These variability threshold limits were defined using the CSDR2 variability statistics of the QUEST RRL stars \citep{vivas2006} discussed in Section \ref{questrrl}. We did not define the CSDR2 variability statistics threshold limits using the catalogue of RRL stars in Stripe 82 \citep{sesar2010} because we later use the latter catalogue to test our efficiency and completeness levels. 

Of the $\sim$ 21,050 stars with the CSDR2 information, 8,351 stars ($\sim$ 40$\%$) passed the CSDR2 additional variability cuts applied in this section. These stars are retained for further analyses.

\subsection{The Analysis of Variance} \label{aov}
We used the Analysis of Variance (AoV; \citealt{schwarzenberg1989}) technique to find the periods of the 8,351 RRL candidates. Using Fourier methods, the AoV technique calculates variances of the light curves using different periods and tries to detect sharp signals \citep{schwarzenberg1989}. Each calculated signal corresponding to a different trial period (in the 0.1--1.1 days range) is then represented in a periodogram (period vs. power). For each light curve, we chose the period with the highest signal. Having low signal reflects a non-periodic behavior for the phased light curve while a large signal reflects a good periodic behavior.

\begin{figure}
\centering
\includegraphics[scale=0.65]{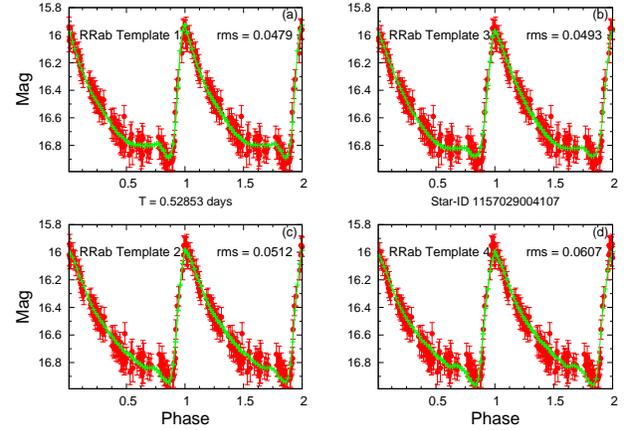}
\caption{Panels (a), (b), (c), and (d) illustrate the four best-fitted templates by the TFM for CSDR2 star-id 1157029004107 (period of 0.52853 days). The phased light curves and the best-fitted templates are plotted in red and green, respectively. All of the four best-fitted templates belong to RRab stars with an $rms$ ranging between 0.0479 and 0.0607. 
\label{bestfour_1157029004107}}
\end{figure}

\subsection{Template Fitting Method} \label{tfm}
After estimating the best-fitted periods using the AoV technique, we used the Template Fitting Method (TFM; \citealt{layden1998,layden1999}) in order to determine the types of our RRL candidates.

In total, the TFM uses a set of 10 different template light curves representing different variable stars. Six of these template light curves are for RRab stars, two for RRc stars, one for W UMa contact binary stars, and one for Algol eclipsing binary stars. The CSDR2 light curves are folded using periods from the AoV technique and are fitted by each of the 10 templates. The $\chi^2_{TFM}$ and $rms$ scatter of each fit are then calculated. Small $\chi^2_{TFM}$ and $rms$ values reflect a good template fit, which in turn reflects the correct type of the star.

In order to insure that this classification process was correct, we visually inspected the four best-fitted templates for each star. The four best-fitted templates for CSDR2 star-id 1157029004107 (period of 0.52853 days) are shown in panels (a), (b), (c), and (d) of Fig. \ref{bestfour_1157029004107}, respectively. The phased light curves and the best-fitted templates are shown in red and green, respectively.  

All of the four best-fitted templates belong to RRab stars with $rms$ ranging between 0.0479 and 0.0607 for the first and fourth best-fitted template, respectively. The asymmetric, steep rise, and slow decrease in brightness of the phased light curves and templates suggest that this is an RRab star with an amplitude of $\sim$ 1.0 mag. This visual inspection was done for all of the RRL candidates.

\section{Results} \label{results}

After applying the AoV and the TFM methods and after the visual inspection of the four best-fitted templates by the TFM, we were able to detect 4,800 RRab stars and 1,571 RRc stars (6,371 RRL stars in total). These are the stars that passed the variability and color cuts and that were well fitted with the templates provided by the TFM. The positions (R.A. and Dec.), CSDR2 mean magnitudes ($Mag$), CSDR2 amplitudes, subtypes, periods, ephemeris (MJD$_{max}$; time at maximum light), and the heliocentric distances ($d_{h}$, see Section \ref{substructure}) of our RRab and RRc stars are found in Table \ref{RRlisttable}. 


\begin{table*}
 \centering
 \begin{minipage}{150mm}
  \caption{The CSDR2 catalogue of RRL stars. Both equatorial J2000.0 R.A. and Dec. are given in decimal degrees. A portion of the table is shown here for guidance regarding its form and content. The table is available in its entirety in the electronic version of the paper, and from the Centre de Donn\'ees Astronomiques de Strasbourg (CDS).}
  \begin{tabular}{@{}lllllllll@{}}
  \hline
   SDSS NAME\footnote{The official SDSS designation for an object where the coordinates are truncated, not rounded, given by the format: JHHMMSS.ss$+$DDMMSS.s} & R.A. & Dec. & $Mag$\footnote{The CSDR2 mean magnitude} & $Amp$\footnote{The CSDR2 amplitude range} & Type & Period\footnote{Period in days} & MJD$_{max}$\footnote{Ephemeris of the stars (time at maximum light)} & $d_{h}$\footnote{Heliocentric distances in kpc}   \\
 \hline
  SDSS J140016.30+155821.4 & 210.0679 & 15.9726 & 17.54 & 1.03 & ab & 0.4860 & 55676.21067 & 27.6\\
  SDSS J170343.25+115155.5 & 255.9302 & 11.8654 & 15.61 & 1.43 & ab & 0.4695 & 54591.39047 & 11.7\\
  SDSS J143301.30+181254.3 & 218.2554 & 18.2150 & 16.32 & 0.93 & ab & 0.4865 & 54228.29835 & 15.9\\
  SDSS J103555.83+382214.0 & 158.9826 & 38.3705 & 16.25 & 1.15 & ab & 0.4875 & 54566.25784 & 14.8\\
  SDSS J153604.59+210746.7 & 234.0191 & 21.1296 & 16.88 & 0.8 & ab & 0.6701 & 53866.34446 & 16.1\\
  SDSS J134644.45+454526.2 & 206.6852 & 45.7572 & 15.76 & 1.2 & ab & 0.4738 & 54138.43831 & 11.8\\
  SDSS J201518.96-124928.8 & 303.8290 & -12.8246 & 16.61 & 0.75 & ab & 0.5559 & 54286.39649 & 15.5\\
  SDSS J172732.63-133844.0 & 261.8859 & -13.6455 & 15.20 & 0.91 & ab & 0.5558 & 53986.47201 & 5.3\\
  SDSS J144412.93+203641.2 & 221.0538 & 20.6114 & 14.37 & 0.47 & c & 0.3479 & 54884.44511 & 7.4\\
  SDSS J145428.85+501007.7 & 223.6202 & 50.1687 & 17.03 & 0.53 & c & 0.3662 & 56126.21641 & 21.2\\
  \hline
\end{tabular}
\label{RRlisttable}
\end{minipage}
\end{table*}

\begin{figure}
\centering
\includegraphics[scale=0.65]{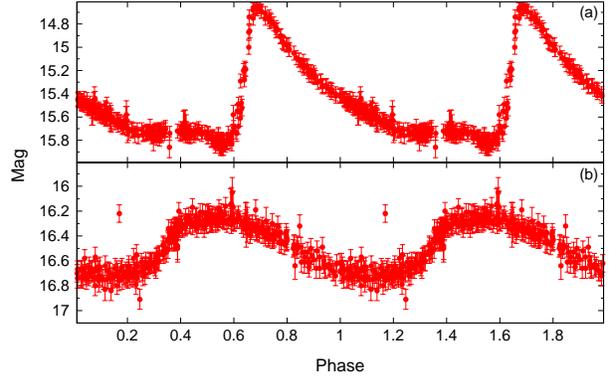}
\caption{Illustration of one of our RRab and RRc stars phased light curves shown in panels (a) and (b), respectively. 
\label{lc_ab_c}}
\end{figure}

The upper and lower panels of Fig. \ref{lc_ab_c} illustrate the phased light curves of one of our RRab and RRc stars, respectively. The asymmetrical shape, steep rise, and slow decrease of the RRab stars' phased light curves have been observed in all of our RRab stars. On the other hand, the more symmetrical phased light curves with relatively smaller amplitudes were observed in all of our RRc stars. The unique phased light curve shape of RRab stars makes it relatively easy to identify them with high efficiency, compared to RRc stars that have phased light curves that can be confused with other types of variable stars (e.g. W UMa, $\delta$ Scuti, and SX Phe stars, see Section \ref{conta}). 

\begin{figure}
\centering
\includegraphics[scale=0.48]{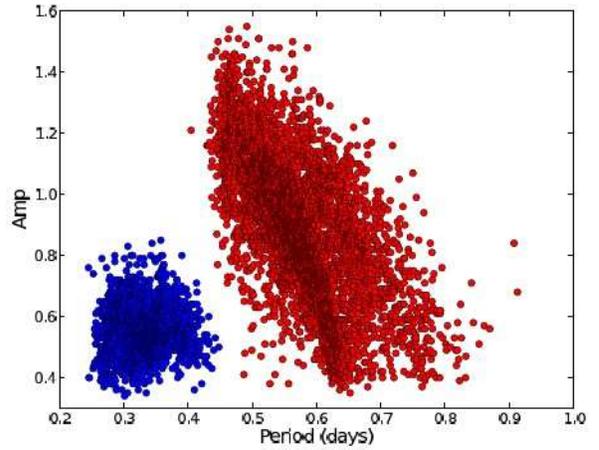}
\caption{The period-amplitude distribution for the RRab (red dots) and RRc (blue dots) stars.
\label{PerVsAmp}}
\end{figure}

The period-amplitude diagram for the RRab (red dots) and RRc (blue dots) stars is shown in Fig. \ref{PerVsAmp}. RRab stars tend to have higher amplitudes and periods than RRc stars, as expected. Fig. \ref{PerVsAmp} also shows that most of the RRab stars are concentrated in a narrow period-amplitude range (periods less than $\sim$ 0.65 days), these stars belong to the Oosterhoff I group (Oo I). Some RRab stars lie toward the longer period region and belong to the Oosterhoff II group (Oo II). The different Oosterhoff groups \citep{oosterhoff1939} depend on the stellar metallicities, horizontal branch morphology, and ages \citep{catelan2009}. RRL stars in the Oo I group are more metal rich compared to RRL stars in the Oo II group. More than 75$\%$ of our RRab stars have periods $\textless 0.65$ days and thus are more likely to belong to the Oo I group. This result agrees well with several studies that have demonstrated that $\ge 70\%$ of the RRab stars in the halo belong to the Oo I group (e.g. \citealt{miceli2008,zinn2014}). Using the CSDR2 light curves, we are planning to conduct a detailed study in the near future about the properties and characteristics of this phenomenon.

\section{Comparison with Stripe 82} \label{comp_s82}
It is important to check how complete, efficient, and reliable our catalogue is. Accordingly, we compare our RRL star discoveries with the catalogue of RRL stars in Stripe 82 \citep{sesar2010}. We also discuss the properties of the RRL stars that we missed.

\subsection{Completeness} \label{complete}

Our catalogue contains 184 RRL stars (137 RRab and 47 RRc stars) in the Stripe 82 area, out of which 177 (136 RRab and 41 RRc stars) are also found in the catalogue of RRL stars in Stripe 82 \citep{sesar2010}. The 7 extra RRL stars are found in our catalogue only. The presence of the extra RRL stars can either prove that our catalog is slightly contaminated by non-RRL stars or that \citeapos{sesar2010} catalog is not 100$\%$ complete.

Our achieved completeness level is distance-dependent because of the magnitude cut ($12.0\textless u \textless19.0$) applied and discussed in Section \ref{ColorBoxes}. All of the common 177 RRL stars in Stripe 82 that are found in our and in \citeapos{sesar2010} catalogue have $d_{h}$ between $\sim$ 4 and $\sim$ 28 kpc. These distances are provided in the latter catalogue. Accordingly, the completeness level will be studied and analyzed in the mentioned distance range.

\citeapos{sesar2010} catalogue contains 337 RRL stars (255 and 82 of type ab and c, respectively) with $d_{h}$ in the 4--28 kpc distance range. Since we recovered 136 and 41 RRab and RRc stars, respectively, our completeness level is then $\sim$ 50$\%$ for RRab and RRc stars.

\subsection{Missed RRL Stars } \label{missed_rr}

With our 50$\%$ completeness level computed in the previous section, we know that we missed $\sim$ 50$\%$ of the RRL stars in Stripe 82 with $d_{h}$ in the 4--28 kpc distance range. These missed stars either did not have CSDR2 information or did not pass the variability cuts in Section \ref{ps1variablity} ($\sigma_{g_{P1}} \textgreater 0.05$ or $\sigma_{r_{P1}} \textgreater 0.05$). 

For example, only $\sim$ 60$\%$ of the RRL candidates that passed the SDSS color and the PS1 variability cuts had CSDR2 data. The remaining missed stars were either not observed with the Catalina Survey, or they are located in crowded regions (e.g. de-blending problems) where they had a small number of clean CSDR2 detections.

Another reason for why we missed 50$\%$ of the RRL stars are the PS1 variability cuts applied in Section \ref{ps1variablity}. Although most of the missed RRL stars were observed $\sim$ 4 times in $g_{P1}$ or $r_{P1}$, the variability of the missed RRL stars did not appear in the PS1 database. RRL stars are short period variable stars and are repeating their cycles between $\sim$ 1 and $\sim$ 5 times a day. Hence, it is likely that some RRL stars were repeatedly observed at the same or a close phase. Nevertheless, we were still able to recover more than 50$\%$ of the RRL stars with the small number of epochs available from the PS1 at the moment. The completeness level will be significantly higher when more PS1 epochs are available. Nevertheless, the variability cuts were necessary to distinguish a possible variable from a non-variable star.

\subsection{Efficiency} \label{s82_efficiency}

Among the 184 RRL stars we found in Stripe 82, 137 and 47 are of type ab and c, respectively. Out of the 137 RRab stars, 136 are classified as RRab stars in \citeapos{sesar2010} catalogue.

Of our 47 RRc stars, 41 are found and classified as RRc stars in the latter catalogue. Assuming that \citeapos{sesar2010} catalogue is complete, our efficiency levels are then $\sim$ 99$\%$ and $\sim$ 87$\%$ for RRab and RRc stars, respectively.

\begin{figure}
\centering
\includegraphics[scale=0.7]{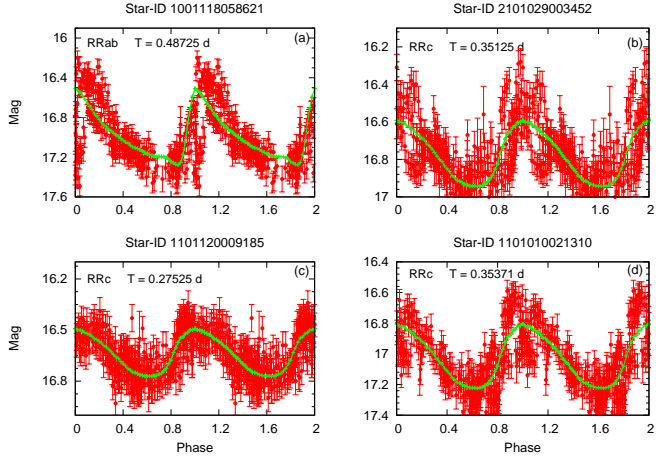}
\caption{The phased light curves of the RRab star and 3 out of the 6 RRc stars that are found in our catalogue but not in \citeapos{sesar2010} Stripe 82 catalogue. The phased light curves and the best-fitted templates from the TFM are plotted in red and green, respectively.
\label{s82missed_rrs}}
\end{figure}

The phased light curves of the extra RRab star and 3 out of the 6 RRc stars are plotted in red in Fig. \ref{s82missed_rrs}. These are the RRL stars in Stripe 82 region that are found in our catalogue but not in \citeapos{sesar2010} catalogue. The best-fitted templates from the TFM are shown in green for stars with the CSDR2 star-ids 1001118058621, 2101029003452, 1101120009185, and 1101010021310 in panels (a), (b), (c), and (d) of Fig. \ref{s82missed_rrs}, respectively. These stars passed the color and variability cuts that are well defined for RRL stars, have been well fitted with RRL star templates by the TFM, and were observed by the CSDR2 between $\sim$ 250 and $\sim$ 380 times. We believe that the phased light curve shown in Fig. \ref{s82missed_rrs}a belong to an RRab star that was missed by \citet{sesar2010}. However, we are less confident about the 6 extra RRc stars that we found in Stripe 82 as their light curves can be confused with other types of variable stars (see Section \ref{conta}). In the worst-case scenario, if we assume that all of the 6 RRc stars are non-RRL stars, our efficiency level would be $\sim$ 87$\%$.

\subsection{Period Testing}

Finally, $\sim$ 95$\%$ of our periods differ on average by only 0.009$\%$ from the periods found by \citet{sesar2010}. The maximum percentage difference was $\sim$ 0.17$\%$. Periods in \citeapos{sesar2010} catalogue were obtained using the Supersmoother routine \citep{reimann1994} which is a smoothing routine that fits data points as a function of phase to a range of frequencies. It uses a running mean or running linear regression on the data points. The small percentage difference between our and \citeapos{sesar2010} periods demonstrates the reliability of our method.

\section{Comparison With the CSDR and the La Silla QUEST catalog of RRL stars} \label{drake_csdr}

\subsection{The CSDR catalog of RRL stars}
In total, there are $\sim$ 14,500 RRab stars found in both the first catalog of RRL stars in the CSDR1 database \citep{drake2013a}, and in its re-analysis study \citep{drake2013b}. These stars were chosen using the Welch-Stetson variability index ($I_{WS}$; \citealt{welch1993}), the Lomb-Scargle periodogram analysis (LS; \citealt{lomb1976,scargle1982}), and the M-Test \citep{kinemuchi2006}. While the $I_{WS}$ measures the variability and tries to separate variable from non-variable stars, the LS looks for periodicity in a specific period range, and the M-Test measures the percentage time spent by the object below the mean magnitude. \citet{drake2013a} used the AoV technique and the Adaptive Fourier Decomposition (AFD) method (G. Torrealba et al., in preparation) to find the periods of their RRab stars.

Of the 14,500 RRab stars found in the CSDR catalogue of RRab stars, $\sim$ 7,500 RRab stars are located in our $d_{h}$ distance range (4--28 kpc) and area (SDSS$\times$PS1$\times$CSDR2 footprint). 

We were able to recover $\sim$ 55$\%$ ($\sim$ 4,150 stars) of the 7,500 RRab stars that are found in the CSDR catalogue of RRL stars \citep{drake2013a,drake2013b}. Missing the remaining $\sim$ 45$\%$ RRab stars was expected as these stars did not show any sign of variability in the PS1 data due to the small number of detections in $g_{P1}$ and $r_{P1}$ that we discussed in Section \ref{missed_rr}. Comparing our periods with periods from the CSDR catalogue of RRL stars for the 4,150 common RRL stars made us trust our results as 99$\%$ of the matched periods had percentage difference less than 0.009$\%$. 

\begin{figure}
\centering
\includegraphics[scale=0.7]{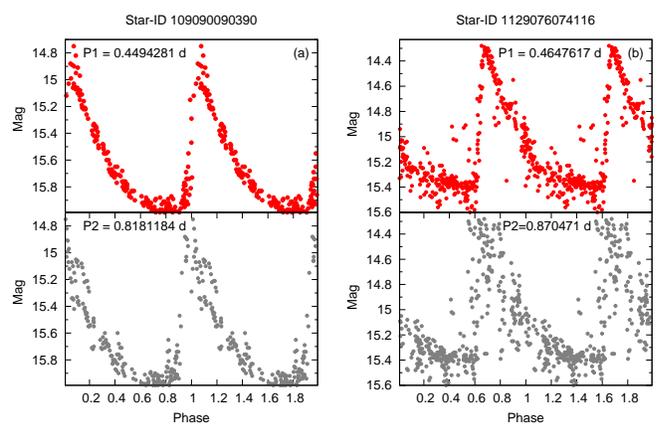}
\caption{Panels (a) and (b) show the phased light curves of CSDR2 star-id 1109090090390 and 1129076074116, respectively. The light curves phased to our periods (P1) and to \citeapos{drake2013a} periods (P2) are shown in red and gray, respectively.
\label{missedper}}
\end{figure}

Panels (a) and (b) of Fig. \ref{missedper} show the phased light curves of CSDR2 star-id 1109090090390 and 1129076074116, respectively. The light curves phased to our periods (P1) and to \citeapos{drake2013a} periods (P2) are shown in red and gray, respectively. Light curves phased in red are less scattered and look more like RRab stars than those plotted with gray implying that our periods are more accurate. We applied the TFM on the phased light curves twice (using P1 and P2). The $rms$ values for the P1-phased light curves were smaller than those corresponding to P2-phased light curves, in both cases. Hence, we trust our claimed periods in the mis-matched cases.

\subsection{La Silla QUEST catalog of RRL stars}

The LSQ catalog of RRL stars \citep{zinn2014} contains 1,372 RRL stars (1,013 RRab and 359 RRc) with $d_{h} \textless 80$ kpc. Around 70$\%$ of the RRab stars found in the LSQ catalog of RRL stars are also found in the CSDR catalogue of RRL stars \citep{drake2013a,drake2013b} that we already compared to our catalog in the previous section. We recovered 355 out of the 1,013 LSQ RRab stars of which 336 are also found in the CSDR catalogue of RRL stars. Only 84 RRc stars are found in both our and LSQ catalog of RRL stars. Most of the remaining LSQ RRL stars that we missed are located beyond our covered distance range ($d_{h} \textgreater$ 28 kpc).

Our catalog contains $\sim$ 230 RRL stars that are not found in the LSQ catalog of RRL stars. \citet{zinn2014} computed their completeness ($\sim$ 70$\%$) level by comparing their catalog with the CSDR catalogue of RRL stars \citep{drake2013a,drake2013b}. We believe that their completeness level is slightly lower than 70$\%$ as we have shown in this study that the CSDR catalogue of RRL stars is not complete. The additional visual inspection that we performed for the light curves of the $\sim$ 230 RRL stars suggests that these are indeed RRL stars.

\section{Newly Discovered RRL Stars} \label{newrrs}

Although \citet{drake2013a, drake2013b} searched for RRab stars in the whole CSDR database, we were still able to find 646 new RRab stars that were missed by them. Of these 646 RRab stars, 19 stars were found also in the LSQ catalog of RRL stars (but not in the CSDR catalogue of RRL stars). Around 572 of our RRab stars are new discoveries. Additionally, we present the discovery of 1,571 RRc stars of which $\sim$ 1,449 are new discoveries. Based on our analyses in Section \ref{comp_s82}, we estimate that only $\sim$ 2$\%$ and $\sim$ 13$\%$ of our RRab and RRc star discoveries are non-RRL stars (contaminant stars). We discuss the nature of our contaminant stars in Section \ref{conta}. 

\subsection{RRab Stars}

We identified 627 additional RRab stars that are neither found in the CSDR \citep{drake2013a,drake2013b} nor in the LSQ catalogs of RRL stars \citep{zinn2014}. The phased light curves, TFM and AoV analyses, and variability statistics of these stars are similar to the other common RRL stars that we and \citet{drake2013a,drake2013b} and \citet{zinn2014} have found.

Out of the 627 newly discovered RRab stars, 55 are found in the GCVS \citep{samus2009}. 42 out of the 55 RRL stars are classified as RRab stars in the GCVS, 12 as RRL stars without giving the sub-type and one as an RRc star.

\begin{figure}
\centering
\includegraphics[scale=0.7]{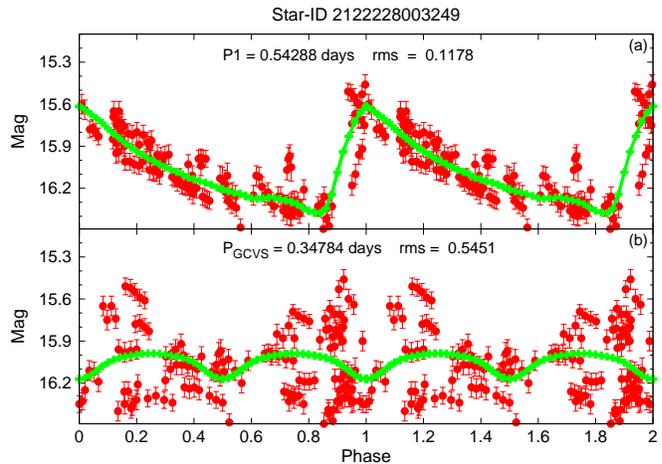}
\caption{A comparison between the two phased light curves (in red) and the best-fitted templates by the TFM (in green) representing our (P1 = 0.54288 days) and the GCVS (P$_{GCVS}$ = 0.34784 d) periods are represented in panels (a) and (b) for CSDR2 star-id 2122228003249. 
\label{missed_gcvs}}
\end{figure}

Our period for the RRc star mentioned earlier (CSDR2 star-id 2122228003249) does not match the period found in the GCVS. The two phased light curves (in red) and the best-fitted templates by the TFM (in green) corresponding to our period (P1 = 0.54288 days) and to the GCVS's period (P$_{GCVS}$ = 0.34784 d) are shown in panels (a) and (b) of Fig. \ref{missed_gcvs}, respectively. The two cycles observed in one phase and the large scatter ($rms$ = 0.5451) around the TFM template in Fig. \ref{missed_gcvs}b suggest that this is not an RRc star, even if we assume that P$_{GCVS}$'s period is correct. However, the RRab-like phased light curve and the small scatter ($rms$ = 0.1178) around the best-fitted RRab template from the TFM in Fig. \ref{missed_gcvs}a indicate that our classification and period are more accurate. 


\subsection{RRc Stars}
Our catalogue contains 1,571 RRc stars, of which 84 are found in the LSQ catalog of RRL stars \citep{zinn2014}. Additionally, only 38 of our RRc stars are found in the GCVS out of which 25 are classified as RRc stars, four as RRL stars without a sub-type, six as RRab stars, and three as stars in eclipsing binary systems (EW stars).

\begin{figure}
\centering
\includegraphics[scale=0.7]{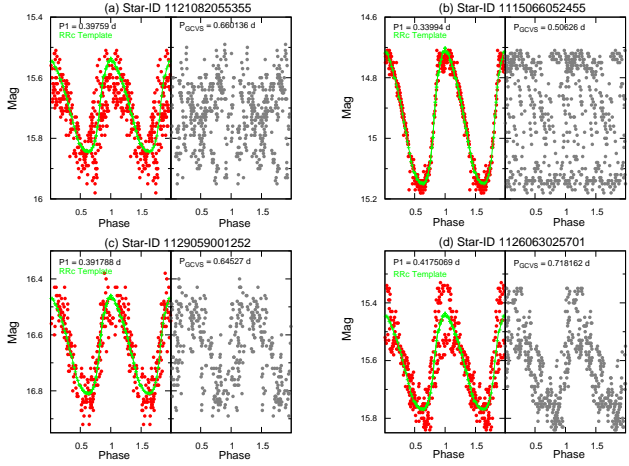}
\caption{The phased light curves corresponding to our (left part of each panel) and to the GCVS (right part of each panel) periods are shown in red and gray, respectively. The green fits indicate the best-fitted templates from the TFM, all corresponding to RRc stars.
\label{missed_rrcs_gcvs}}
\end{figure}

Four out of the six RRc stars that were classified as RRab stars have periods in the GCVS. The phased light curves corresponding to our (left part of each panel) and to the GCVS (right part of each panel) periods (P$_{GCVS}$) are shown in red and gray in Fig. \ref{missed_rrcs_gcvs}, respectively. The green fits indicate the best-fitted templates from the TFM, all corresponding to RRc stars. It is clear that folding the light curves to the P$_{GCVS}$ periods (in gray) did not produce periodic signals in the panels (a), (b), and (c) of the latter figure. On the other hand, periodic signals (in red) and well fitted RRc templates (in green) were observed when folding the light curves to our periods, indicating that these are indeed RRc stars. 

Although periodic signals are observed when folding the data to our and to the P$_{GCVS}$ period in Fig. \ref{missed_rrcs_gcvs}d, we believe that our period and classification are more accurate because the AoV and TFM analyses are based on 274 CSDR2 epochs for this star.

\subsection{Contaminant Stars} \label{conta}

It is not surprising that some of our RRc stars are contaminated by other type of stars (e.g. W UMa stars, $\delta$ Scuti, etc.) as these stars have moderately symmetric light curves (e.g. sinusoidal) and share the same period range. However, we used color cuts that are well characterized for RRL stars. Additionally, each light curve was fitted to both RRc and W UMa templates by the TFM. The fits worked better for the RRc templates for all of our RRc stars.

After comparing our RRc star discoveries with the GCVS in the previous section, and with Stripe 82 in Section \ref{s82_efficiency}, we believe that the RRc contamination level by eclipsing binaries is less than $\sim$15$\%$, and is caused mainly due to W UMa stars.

Finally, $\delta$ Scuti and SX Phe stars have colors similar to the colors of RRL stars \citep{palaversa2013} and have periods less than $\sim$ 0.3 days\footnote{http://www.aavso.org/types-variables}. Because almost half of our RRc stars have periods less than 0.3 days, $\sim$ half of our RRc stars are prone to contamination from $\delta$ Scuti and SX Phe stars.

\section{Halo Sub-Structure} \label{substructure}

In this section, we check whether the completeness and efficiency levels of our catalogue are good enough to detect previously known and possibly new halo overdensities. First, we derive the $d_{h}$ of the RRL stars in our catalogue using Equation \ref{Heli_Distx}:

\begin{equation} \label{Heli_Distx}
d_{h} = 10^{(\langle V_{0} \rangle-M_{v}+5)/5}
\end{equation} 

\noindent
where $\langle V_{0} \rangle$ magnitudes are calculated using Equation \ref{VMag} which was adopted from \citet{ivezic2005}.


\begin{equation} \label{VMag}
\langle V_{0} \rangle = r - 2.06(g-r) + 0.355
\end{equation} 

\noindent
where the $g$ and $r$ SDSS magnitudes were corrected for the line-of-sight interstellar extinction using the recalibration of \citeapos{schlegel1998} dust map by \citet{schlafly2011}. This equation corrects a bias in single-epoch SDSS measurements due to the unknown phase and introduces a minimal rms scatter of 0.12 mag.

Like \citet{sesar2010}, we adopt $\langle M_{V} \rangle$ = 0.60 mag for the absolute magnitude of RRL stars; a value that was calculated by \citep{cacciari2003} using Equation \ref{MvFe}.

\begin{align}
\label{MvFe}
M_{V} = (0.23 \pm 0.04)\mathrm{[Fe/H]}+(0.93 \pm 0.12)
\end{align}

\noindent
where the mean halo metallicity of [Fe/H] = $-1.5$ $\pm$ 0.32 dex is used \citep{ivezic2008}. Adopting [Fe/H] = $-1.5$ dex introduces $rms_{M_v}$ of $\sim$ 0.1 mag because of the actual dispersion of [Fe/H] and their corresponding uncertainties. Taking the uncertainties of [Fe/H], $\langle V_{0} \rangle$, and ${M_v}$ into account, $d_{h}$ is calculated with at least $\sim$ 7\% fractional error.

\begin{figure}
\centering
\includegraphics[scale=0.4]{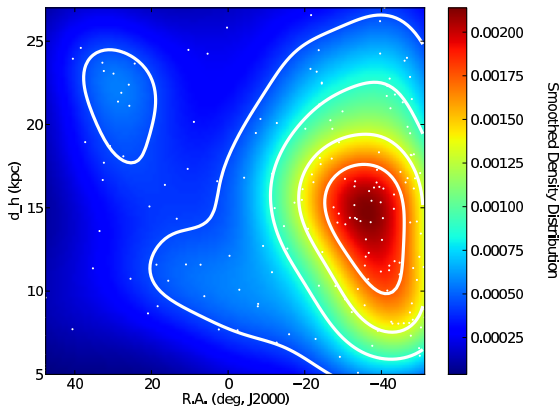}
\caption{The number density distribution of the 184 RRL stars found in our catalogue in the Stripe 82 area. The density of the points that is accentuated by the white contours is shown in scaled density levels. The smoothed surface regions with high and low numbers of stars are represented in red and dark blue, respectively. The Hercules\--Aquila cloud halo structure appears at R.A. $\sim$ $-40\,^{\circ}$ and $d_{h}$ in the 8 kpc to 24 kpc distance range while the arm of the Sagittarius dwarf spheroidal (dSph) tidal stream appears at R.A. $\sim$ $ 30\,^{\circ}$ and $d_{h}$ $\sim$ 23 kpc. Negative values of R.A. were used for better visualization only (R.A. = R.A. + 360$^{\circ}$ when R.A. $\textless$ 0$^{\circ}$).
\label{Stripe82_south}}
\end{figure}

To test if our completeness and efficiency levels are good enough to detect substructures (and to possibly find new ones), we plotted the number density distribution of the 184 RRL stars we found in Stripe 82 in Fig. \ref{Stripe82_south}. Assuming that \citeapos{sesar2010} catalogue is 100$\%$ efficient and complete implies that the contamination level of our 184 RRL stars is $\sim$ 4$\%$. The density of the points that is accentuated by the white contours is shown in scaled density levels. The smoothed surface regions with high and low numbers of stars are represented in red and dark blue, respectively. 

The Hercules\--Aquila cloud \citep{belokurov2007} halo substructure appears at R.A.\footnote{Add 360\,$^{\circ}$ to obtain the correct values of R.A. when R.A. $\textless$ 0$^{\circ}$. Negative values of R.A. were used for better visualization only.} $\sim$ $-40\,^{\circ}$ and spans $d_{h}$ in the 8 to 24 kpc distance range while the arm of Sagittarius dwarf spheroidal (dSph) tidal stream \citep{majewski2003,majewski2010} appears at R.A. $\sim$ $ 30\,^{\circ}$ and $d_{h}$ $\sim$ 23 kpc.

\begin{figure}
\centering
\includegraphics[scale=0.4]{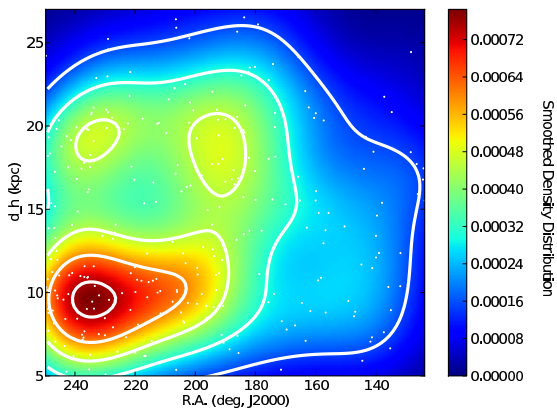}
\caption{The number density distribution of the RRL stars found in the Northern Galactic hemisphere section of the celestial equator ($-1.25\,^{\circ} \textless$ Dec. $\textless 1.25\,^{\circ}$). Two main structures are detected: the Virgo overdensity at R.A. $\sim$ $190\,^{\circ}$ and $d_{h}$ $\sim$ 19 kpc, the Hercules\--Aquila cloud at R.A. $\sim$ $240\,^{\circ}$ and $d_{h}$ $\sim$ 10 kpc. 
\label{dec1.25_north}}
\end{figure}

Our RRL stars that are found in the Northern Galactic hemisphere section of the celestial equator ($-1.25\,^{\circ} \textless$ Dec. $\textless 1.25\,^{\circ}$) are plotted in Fig. \ref{dec1.25_north}. The Virgo overdensity \citep{vivas2001} was detected at R.A. $\sim$ $190\,^{\circ}$ and $d_{h}$ $\sim$ 19 kpc while the Hercules\--Aquila cloud at R.A. $\sim$ $240\,^{\circ}$ and $d_{h}$ $\sim$ 10 kpc.

Some parts of the well-defined narrow tidal tails of the extended, low-concentration globular cluster Palomar 5 (Pal 5; \citealt{odenkirche2001, odenkirche2003,grillmair2006}) overlap in projection with the clump appearing at R.A. $\sim$ $235\,^{\circ}$ and $d_{h}$ $\sim$ 20 kpc (see Fig. \ref{dec1.25_north}). Pal 5 is a faint halo cluster that is currently undergoing tidal disruption due to disc shocks \citep{dehnen2004}. This globular cluster has a very sparsely populated red giant branch and horizontal branch \citep{odenkirche2003}, with only very few RRL stars (five, see \citealt{vivas2006}). Two additional RRL stars have been suggested to be associated with Pal 5's tidal tails \citep{vivas2006}. Although the clump appearing at R.A. $\sim$ $235\,^{\circ}$ and $d_{h}$ $\sim$ 20 kpc can indeed be associated with Pal 5, we do not confirm this association because of the small number of RRL stars (5--7) associated with Pal 5 in addition to possible stars that are contaminating our catalog. We suggest radial velocity studies of our RRL stars that belong to this clump in order confirm the association of this clump to Pal 5.

\section{Summary} \label{summary}

We have combined data from different sky surveys (the SDSS, the PS1, and the Catalina Survey) to look for RRL stars in the Milky Way halo. The search resulted in the discovery of 6,371 RRL stars (4,800 RRab and 1,571 RRc) distributed around 14,000 deg$^2$ of the sky and with $d_{h}$ in the 4--28 kpc distance range. Around 2,021 ($\sim$ 572 RRab and 1,449 RRc) of these stars are new discoveries. 

In this paper, RRL stars were discovered using the SDSS color and the PS1 variability cuts in Section \ref{identifyrrl}. We define the threshold limits of these cuts using the QUEST catalogue of RRL stars \citep{vivas2006} rather than using the catalogue of RRL stars in Stripe 82 \citep{sesar2010} as we use the latter catalogue to test the efficiency and completeness levels of our method.

Additional variability cuts were applied and light curves were plotted using the CSDR2 multi-epoch data. Periods were obtained using the AoV technique while the classification process was done by the TFM and by visual inspection. The comparison of our RRL star discoveries with the RRL stars in Stripe 82 from the SDSS shows that our completeness levels are $\sim$ 50$\%$ for RRab and RRc stars and that our efficiency levels are $\sim$ 99$\%$ and $\sim$ 87$\%$ for RRab and RRc stars, respectively.

Additional comparison of our RRL star discoveries with the GCVS, the LSQ catalog of RRL stars, and the 14,500 RRab stars found previously in the Catalina Survey \citep{drake2009,drake2013b} suggests the reliability of our method. Additionally, the Virgo overdensity, Hercules\--Aquila cloud, and Sagittarius stream were recovered after plotting the number density distribution of our RRL stars in the Stripe 82 and Northern Galactic hemisphere areas. This indicates that our method is capable of identifying halo overdensities. In a forthcoming paper, we will present a more detailed analysis of halo substructure as traced by RRL stars.

\section*{Acknowledgments}

We thank the referee for comments and constructive suggestions that helped to improve the manuscript. We thank E. Bernard, J. Vanderplas, S. Duffau, and A. Huxor for helpful discussion that improved the quality of this paper. M.A., E.K.G., and N.F.M acknowledge support by the Collaborative Research Center ``The Milky Way System" (SFB 881, subproject A3) of the German Research Foundation (DFG). The Pan-STARRS1 Surveys (PS1) have been made possible through contributions of the Institute for Astronomy, the University of Hawaii, the Pan-STARRS Project Office, the Max-Planck Society and its participating institutes, the Max Planck Institute for Astronomy, Heidelberg and the Max Planck Institute for Extraterrestrial Physics, Garching, The Johns Hopkins University, Durham University, the University of Edinburgh, Queen's University Belfast, the Harvard-Smithsonian Center for Astrophysics, the Las Cumbres Observatory Global Telescope Network Incorporated, the National Central University of Taiwan, the Space Telescope Science Institute, the National Aeronautics and Space Administration under Grant No. NNX08AR22G issued through the Planetary Science Division of the NASA Science Mission Directorate, the National Science Foundation under Grant No. AST-1238877, the University of Maryland, and Eotvos Lorand University (ELTE). Funding for SDSS-III has been provided by the Alfred P. Sloan Foundation, the Participating Institutions, the National Science Foundation, and the US Department of Energy. The SDSS-III Web site is http://www.sdss3.org/. SDSS-III is managed by the Astrophysical Research Consortium for the Participating Institutions of the SDSS-III Collaboration including the University of Arizona, the Brazilian Participation Group, Brookhaven National Laboratory, University of Cambridge, University of Florida, the French Participation Group, the German Participation Group, the Instituto de Astrofisica de Canarias, the Michigan State/Notre Dame/JINA Participation Group, Johns Hopkins University, Lawrence Berkeley National Laboratory, Max Planck Institute for Astrophysics, New Mexico State University, New York University, Ohio State University, Pennsylvania State University, University of Portsmouth, Princeton University, the Spanish Participation Group, University of Tokyo, University of Utah, Vanderbilt University, University of Virginia, University of Washington, and Yale University. CRTS is supported by the U.S. National Science Foundation under grants AST-0909182 and CNS-0540369. The work at Caltech was supported in part by the NASA Fermi grant 08-FERMI08-0025 and by the Ajax Foundation. The CSS survey is funded by the National Aeronautics and Space Administration under grant No. NNG05GF22G issued through the Science Mission Directorate Near-Earth Objects Observations Program.

\label{lastpage}
\end{document}